\documentclass[a4paper,11pt]{article}
\usepackage{pos}

\title{On the overflowing of cosmic rays from galaxies and the expansion of cosmic matter}

\author*[a]{Antonio Codino}

\affiliation[a]{University of Perugia and INFN,\\
  Via A. Pascoli, Perugia, Italy}

\emailAdd{antonio.codino@pg.infn.it}

\abstract{Particles of the cosmic radiation, electrons and nuclei,
transport a dominant positive electric charge. A tiny fraction of
these particles of extremely high energies in favorable conditions
overflow from galaxies. The overflowing of positively charged cosmic
nuclei into the intergalactic space uncovers an equal amount of
negative charge in the parent galaxy. Negative charge is mainly
stored by quiescent electrons. After adequate particle propagation
neither the negative electric charge located in the galaxies nor the
positive electric charge of the overflowed cosmic nuclei can be
neutralized due to the enormous distances.

\quad In several ways it is proved that the total electric charge
retained by clusters of galaxies after an appropriate time interval
generate a repulsive force between clusters which overwhelms
gravity. After a few billions years of electrostatic repulsion,
peripheral clusters attain relativistic velocities and their mutual
distances increase accordingly. Several facts suggest that the
expansion of the universe, as determined by optical observations
since a century, has been caused by the electrostatic repulsion of
the positively charged cosmic nuclei overflowed from galaxy
clusters.}

\FullConference{37$^{\rm{th}}$ International Cosmic Ray Conference (ICRC 2021)\\
		July 12th -- 23rd, 2021\\
		Online -- Berlin, Germany}

\begin{document}
\maketitle

\section{Introduction}
\quad It is a fact known since 1961 that the dominant fraction of
cosmic radiation, about 98 per cent of the total flux above 10-20
$GeV$ consists of atomic nuclei which all have positive electric
charge. An interesting facet of this fact is that the propagation
volume of cosmic rays is permeated by a dominant fraction of
positive electric charge and, consequently,  cosmic-ray sources
retain negative electric charge in order to preserve charge
neutrality on a suitable scale length.

\quad  A second basic fact is that the nearby universe consists of
many spiral galaxies whose cosmic-ray properties have to be
similar to those measured in the $Milky$ $Way$ $Galaxy$. These
properties have been measured through synchrotron radiation
(hereafter curvature radiation) of cosmic electrons in the 10-$100$
$cm$ radio band for more than $100$ spiral galaxies. The fraction of
spiral galaxies in galaxy clusters in the nearby universe is
typically in the range 0.30-0.35 the rest being ellipticals and
irregulars.

\quad This study posits that 
disk galaxies, 
elliptical galaxies and irregular galaxies i.e. all galaxy
categories, store cosmic rays. Support to this prerequisite comes
from radio astronomy data but also from gamma-ray observations.
Gamma rays from proton interactions via neutral pion decays have
been observed in the $Large$ $Magellanic$ $Cloud$,
Andromeda and  $M33$ galaxy. Moreover, a tight correlation
between gamma-ray emissivity and radio emissivity (curvature
emission) in galaxies has been evidenced thereby testifying the presence of cosmic rays.

\quad Quite recently in a long scientific document the electrostatic acceleration of cosmic rays
has been proposed \cite{codino20book}. Following this work disk galaxies retain in their
disks a negative electric charge of about $2$-$3$ $\times 10^{31}$
$C$, order of magnitude. This conclusion entails radical changes in many areas of all macroscopic sciences as demonstrated elsewhere \cite{codino21book}. 

\section{Equilibrium between gravitational and electrostatic forces}
\quad  Consider two spherical objects of mass $m$ and $M$, placed at
the distance  $r$ (fig. 1). Assume further that the two spheres have
the electric charges  $q$ and $Q$, respectively, both either
positive or negative. The charge distributions in the two objects
are such that develop the a point-like repulsive force.  There
exists a unique ratio between the electric charge trapped on the
objects and the masses $m$ and $M$ for which, whatever is the
distance $r$ arbitrarily large or small, the force between the two
spheres is balanced, e. g. is zero. This is the balancing condition
featured by:

\parskip=0.8truecm
$$ G {m M \over r^2 } = {1 \over 4 \pi \epsilon_0 }{q Q \over r^2 }  \eqno (1)$$
\vskip 0.8truecm

\quad  The charge-mass ratio in the balancing condition is :
 ${[(q/m)(Q/M)]}$ = ${ 4 \pi \epsilon_0 G }$.

\quad If, for sake of simplicity,  $m$=$M$ and $q$=$Q$ is imposed,
then the relation :
 \parskip=0.7truecm
 $$ q/m = ( { 4 \pi \epsilon_0  G})^{1 / 2} \eqno (2)$$
 \parskip=0.7truecm
\quad  holds. The $q$/$m$ ratio is   $8.617508609 \times 10^{-11}$
$C$/$kg$ and will be denoted by
 $(q/m)_b$ or $q_b$/$m_b$ where $b$ on foot designates balancing.

\quad  The equality between electrostatic and gravitational force
expressed by the equations  (1) and (2) is a mere definition, devoid
of any subtlety, with the arbitrary conditions $m$=$M$ and $q$=Q.
The amount of electric charge $Q$ stored in a generic macroscopic
body of mass  $m$ can be also expressed by the dimensionless
variable \quad $\zeta$ \quad which adopts as unit charge $q_b$. By
definition it is set : $\zeta$ $\equiv$ $Q$/$q_b$ \quad being $q_b$
that particular amount of electric charge residing in the material
body of mass $m$ such as the electrostatic repulsive force $f_e$ with another
arbitrary material charged body equalizes the attractive
gravitational force $f_g$. 

\begin{figure}
\begin{center}
  \includegraphics[width=8cm]{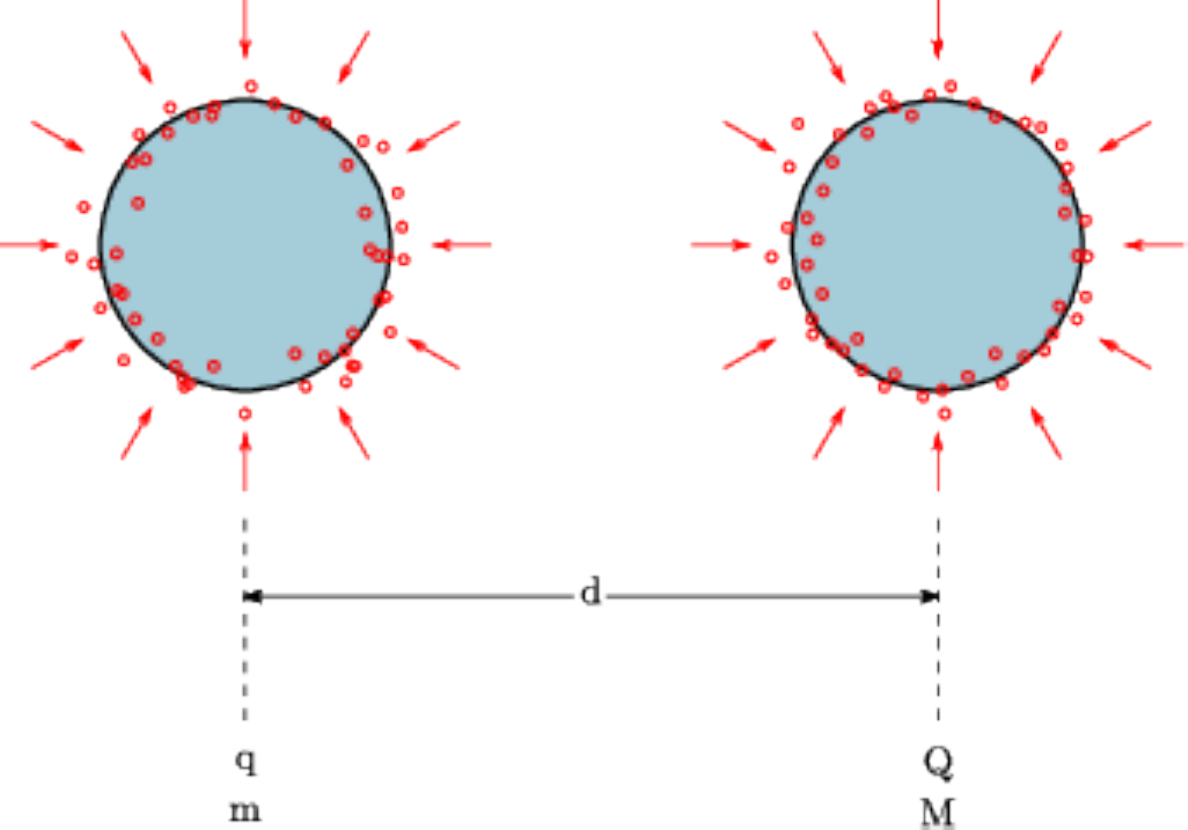}
    \caption{Competition between
gravitational and electrostatic forces of two imaginary material
bodies of mass $m$ and $M$ and electric charge $q$ and $Q$ placed at
the arbitrary distance $d$. Red dots represent cosmic nuclei coming
from the exterior dumped on the outer surfaces of the masses $m$ and
$M$ as qualitatively shown in the drawing. With no positive electric
charge both bodies would precipitate one another by gravity. Instead, with an extremely 
small amount of electric charge relative on the masses $m$ and $M$,
both bodies would repel. 
The particular amount of electric charge with q = Q which maintain
equilibrium of $m$ and $M$
is designated in this work by
$q_b$, and referred to as the balancing charge or equilibrium
charge.}
\end{center}
\end{figure}

\section{Electrostatic repulsion between pairs of rich clusters}
\quad  Let be $r_c$ the typical radius of a rich galaxy cluster of
mass $M_c$ containing $N_g$ galaxies and $R_c$ the maximum extent of
the positive charged halo of cosmic nuclei overflowed from clusters. Moreover,  let be $d_c$
the typical distance between rich galaxy clusters in the nearby Universe. The nominal negative charge stored in the disks of disk galaxies $Q_{w}$ is 2.585252 $\times$ 10$^{31} C$ (see Chap. 5 of ref. \cite{codino20book}). This charge amount is ascribed to the cores (i.e. inner region) of other galaxies composing rich clusters.

\quad Here it is desired to evidence that the electrostatic forces
responsible of cosmic expansion around the Earth in the range 1-10
$Gpc$ or (0.3-3) $\times$ 10$^{26}$ m are generated by the electric charges stored in galaxy
clusters and not by charges in isolated or sparse galaxies. This
assertion is inferred only by the parameters: $r_c$, $N_g$, $d_c$,
$M_c$ and $R_c$ of rich galaxy clusters. For the numerical estimates to follow the calculation
assumptions are : the nominal mass of a generic rich cluster, \quad
$M_c$ = $7.5 \times 10^{47} g$ resulting from the adopted values,
$M_c$ = $N_g$ $\times$ $m_g$ = $2500\times$$3\times$ $10^{44}$ $g$ =
$7.5 \times$$10^{47}$ $g$, r$_c$= 3 Mpc, d$_c$= 60 Mpc and R$_c$=150 Mpc justified elsewhere \cite{codino21book}.
\quad
A useful electric charge unit for galaxy clusters, a dimensionless quantity, is $\zeta_c \equiv$ $Q^c$/ $Q^c_{b}$ where $Q^c_{b}$ = $N_g$$Q_w$ = $6.46313$ $\times
10^{34}$ $C$ is the balancing charge and $Q^c$ the total electric charge within the radius $r_c$ of the cluster.


\quad The phenomenon of electrostatic repulsion between two rich of
clusters is qualitatively illustrated in fig. 2. Let us consider a
rich spherical cluster denoted by $A$ with center $\Omega_A$ which
has the characteristic parameters previously mentioned and two concentric spheres with common
origin $\Omega_A$ and radii $r_c$ and $R_c$ as shown in fig. 2. Let
$\xi_A$ be the electric charge fraction of the cosmic rays stored in
the sphere of radius $r_c$ while the rest of the charge, (1 -
$\xi_A$) is contained in a spherical shell comprised between $r_c$
and $R_c$. The positive charge of cosmic rays within $r_c$ in the
volume \quad ($4$/$3$) $\pi$ $r_c^3$ \quad is:\quad $Q^A_{cr}$($r_c$) = $N_g$
$Q_{cr}$($r_c$/$R_c$)$^2$ = $N_g$ $2.5852 \times 10^{31}$ $C$. The
total electric charge, which is the sum of the positive $Q^A_{cr}$($r_c$) and negative charges $Q_w$ in the spherical volume of radius $r_c$ is : \quad
$Q^A_{e}$ = $Q_w^A$ + $Q^A_{cr}$($r_c$) = -$Q^A_{cr}$ +
$Q^A_{cr}$($r_c$) = -$6.460546 \times 10^{34}$ $C$ \quad  called
here also effective charge of the cluster $A$, denoted
 \quad $Q^A_e$ (e on foot for effective). The electric charge conservation implies:
$Q^A_w$ = $Q^A_{cr}$ ( $1$- $\xi_A$). Numerically it is : $\xi_A$ =
$4.0 \times 10^{-4}$ and  ( $1$- $\xi_A$) = $99.9599 \times
10^{-2}$. The charge to be taken into account in order to compute
the global electrostatic force on the cluster
 $A$ is $Q^A_e$ i. e. the effective charge of the cluster A exerting a repulsive force on the cluster B. 

 \quad  Let designate by $B$ a rich spherical cluster
placed at distance  $d$ from the cluster  $A$ (fig. 2). For
simplicity the cluster $B$ has the  same characteristic features of
the cluster $A$, namely, $\xi^A$ = $\xi^B$, $Q^A_{cr}$ = $Q^B_{cr}$
and $Q^A_{e}$ = $Q^B_{e}$. Let be $\vec E^A$($r$) the electric field
generated by the cluster  $A$ at the generic distance expressed by
the sum of two terms: \quad $\vec E^A$($r$) = $\vec E^A_-$($r$) +
$\vec E^A_+(r)$ \quad where $\vec E^A_-$($r$) is the electric field
generated by the negative charge \quad $Q^A_{e}$ retained within the
radius  $r_c$ of the cluster  $A$ and $\vec E^A_+$($r$) the electric
field generated by the positive charge contained in the spherical
shell between $r_c$ (see fig. 2). The repulsive force $ \vec
f^{AB}(d)$ at the distance $d$ exerted by the cluster $A$ on  $B$ is
given by :

\vskip 0.1truecm
 $$ \vec f^{AB}(d) = {Q^B_{e}\left[ \vec E^A_+ (d) - \vec E^A_- (d) \right]}
 \eqno (3)$$
 \vskip 0.1truecm

\quad where  $Q^B_{e}$ is the effective electric charge of the
cluster $B$. The positive electric charge carried by cosmic nuclei
residing in the spherical shell between $d$ and $R_c$ centered at the origin 
$\Omega_A$ does not exert any force due to the assumed spherical
symmetry of the electrostatic structure under examination. In these
conditions it is inevitable that a repulsive force sets on between
the cluster $A$ and $B$ because: \quad $\vec E^A_-(r)$ $
> $ $\vec E^A_+(r)$. In fact, the field
intensity  $\vec E^A_-(r)$ beyond $r_c$ is:

\vskip 0.1truecm
 $$ E^A_- (r) = \zeta_c  Q^A_{e}/( 4 \pi \epsilon_0 r^2)
 = 5.808 \times 10^{44} \zeta_c /r^2 = 0.610  \zeta_c /r^2 (Mpc) \quad V/m
 \eqno (4)$$
 \vskip 0.1truecm  while the intensity of the field $\vec E^A_+ (r)$
in the range $r_c$ $<$  $r$ $\leq$ $R_c$ is : \vskip 0.1truecm
 $$ E^A_+ (r) = ({ k /2 \epsilon_0)  \zeta_c /(1. - r_c^2/r^2)} =
 ({ 2.704 \times 10^{-5})  \zeta_c /(1. - r_c^2/r^2)}   \quad V/m
 \eqno (5)$$
 \vskip 0.1truecm

\quad with $k$= $4.800$ $\times 10^{-16}$ $C$/$m^2$. The strength of
the field $ \vec E^A_+$ expressed by equation (5) derives from the
 radial profile $\rho_c(r)$ of the positive charge transported by
cosmic nuclei $Q^A_{cr}$ beyond $r_c$. For example, with $\zeta_c$ =
$150$ and $r$ = $d_c$ = $60$ $Mpc$ it is : $E^A_+(d_c)$ = $0.4056 $
$\times 10^{-2}$ $V$/$m$ while $ E^A_-(d_c)$
 = $2.541 \times 10^{-2}$ $V$/$m$,  so that the intensity $ E^A$($d_c$)
is $2.135$ $\times 10^{-2}$ $V$/$m$.

\begin{figure}
\begin{center}
  \includegraphics[width=10cm]{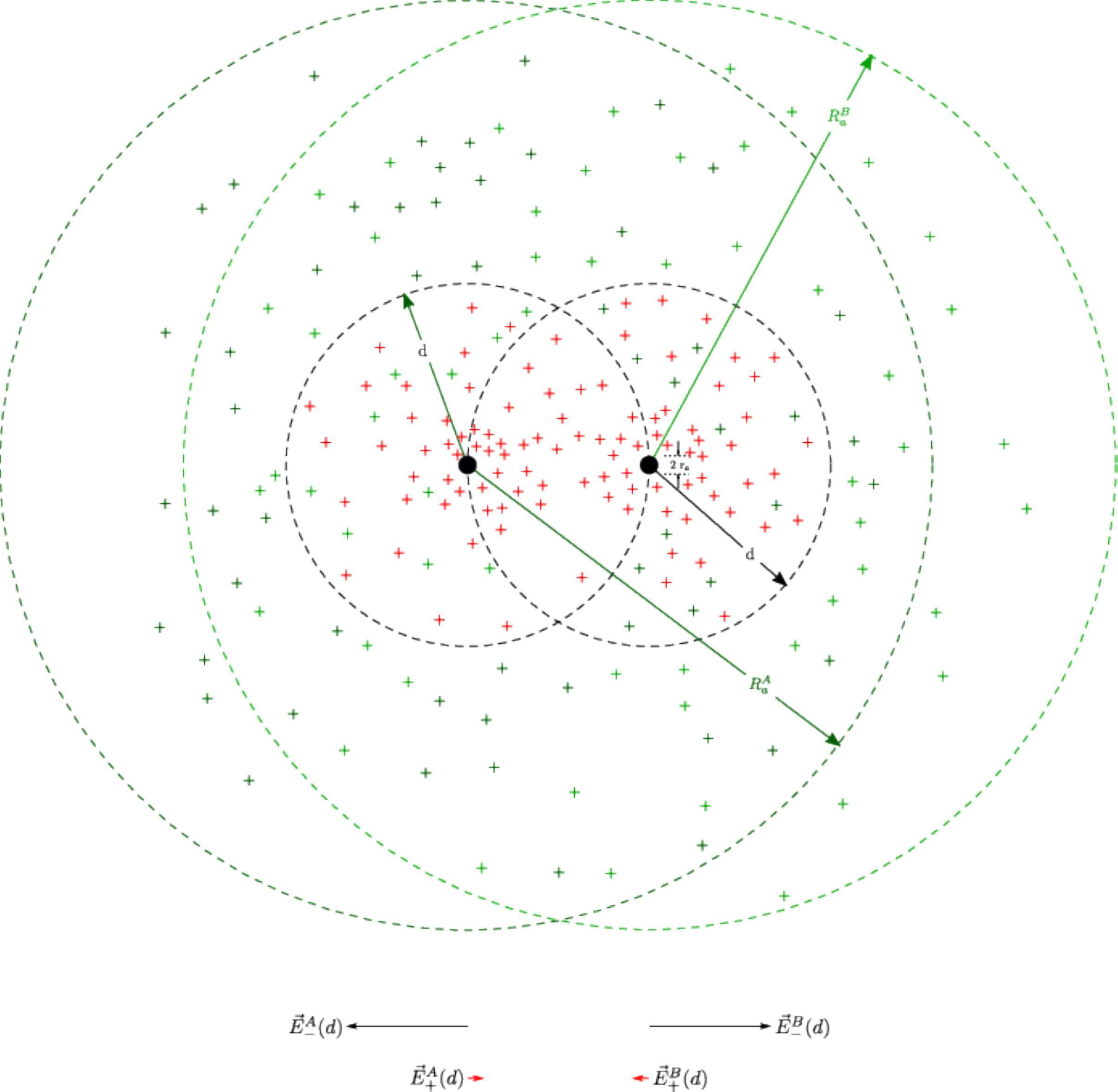}
  \caption{Two galaxy cluster $A$ and $B$ (black
dots) set at the distance $d_c$ = $60$ $Mpc$ with two positive
charged mantels (spherical shells) of radii
 $R^A_a$ and $R^B_b$. The fraction of electric charge that does not affect
the repulsive force between between clusters $A$ and $B$ is
represented by red crosses (dark green for the cluster $A$ and light green for the cluster $B$). The positive electric charge
(red crosses) in the spherical shell between $r_c$ and $d_c$
generates the fields $\vec E_+^A$($d_c$) and $\vec E_+^B$($d_c$)
which exert the attractive force. The negative charge contained
within the radii $r_c$ of the two clusters generate the fields $\vec
E_-^A$($d_c$) and $\vec E_-^B$($d_c$) which exert the repulsive
force. The aforementioned fields, obtained by putting $r_c$ = $3$
$Mpc$, $R_c$ = $150$ $Mpc$ and $\zeta_c$ = $150$, are represented by
vectors in the bottom of the figure.}
\end{center}
\end{figure}

\quad The corresponding electrostatic potential $V^A_e(r)$ of the
cluster $A$ at any arbitrary point at distance $r$ from the origin
$\Omega$ with $r_c$ $\leq$ $r$ $\leq$ $R_c$ is (see Chap. 5 ref. \cite{codino20book}): \vskip 0.1truecm
$$  V^A_e(r) =
{  \zeta_c  Q^c_{b} \over 4 \pi \epsilon_0}  (- {1 \over r} -  { r
\over R_c^2} + {2 \over R_c}) = { \zeta_c N_g  Q_{b} \over 4 \pi
\epsilon_0} ( -{1 \over r} -  { r \over R_c^2} + {2 \over R_c} )
\eqno (6)$$ \vskip 0.1truecm

\quad For example, for $r$ = $r_c$ = $3$ $Mpc$ it is,  $V_c$(3) =
$-2.35\times 10^{27}$ $V$ while for  $r$ = $60$ $Mpc$ it is,
$V_c$(60) = $-3.92\times 10^{26}$ $V$.

\quad From these results it is quickly realized that the
electrostatic field of the cluster $\vec E_c$ at distances of some
dozens of $megaparsec$, is by far greater than that of isolated
galaxies. Indeed, the electric field of sparse galaxies displaying
spherical symmetry $\vec E_g$ vanishes at large distance from the
galactic center as the typical value $R_g$ (for example, $R_g$ =
$0.3$ $Mpc$) is much less that the typical intercluster distance
$d_c$ (for example, $60$ $Mpc$). The electric field intensities of
sparse galaxies devoid of spherical symmetry at large distances $r$
from the origin $\Omega$, fall more rapidly than those having a
quadratic dependence as they depend on multipole terms of the charge distribution. Hence, the
field strength $E_g$, though not zero, it is certainly negligible
relative to the intensities of the fields $E^A$($r_c$) and
$E^B$($r_c$).

\quad The conclusion is
that the electrostatic fields of galaxy clusters dominate the
expansion of the cosmic matter.


\section{Velocities of galaxy clusters and intensities of the electric fields}
\quad In the non relativistic approximation, the acceleration $\vec
a_c$ acquired  by the cluster $B$ subject to the field $\vec
E^A$($d_c$) with $d_c$ = $60$ $Mpc$ as sketched in fig. 2 is given
by:

$$ \vec a_c =  (Q^B_{e}/M_c^B ) \zeta_c   \vec E^A(d_c) =
( { 4 \pi \epsilon_0  G})^{1 / 2}  \zeta_c   \vec E^A(d_c) = 2.7
 \times  10^{-10} ms^{-2}   \eqno (8)$$
\vskip 0.1truecm being  $\vec E^A$($d_c$) = $2.135 \times 10^{-2}$
$V$/$m$, $\zeta_c$ = $150$ and $M^B_c$ = $7.5 \times 10^{44}$ $Kg$.

\quad  An order-of-magnitude estimate of the velocity of the generic
cluster residing in an ambient crowded with galaxy clusters (unlike
that just considered of only two clusters) where $\overline a_c$
represents an average, suitable, constant value of the acceleration
in the given temporal span $t$. For instance, with $\overline a_c$ =
$2.7 \times 10^{-10}$ $ms^{-2}$ in the time $t$ of one billion years
($3.155 \times 10^{16}$ sec), being, \quad  $v$($t$) = $\overline
a_c$ $t$, the velocity reached by the generic galaxy cluster is
$8700$ $Km$/$s$. The straight line segment traveled  in the same
time interval \quad ($1$/$2$) $\overline a_c$ $t^2$ \quad is $4.2$
$Mpc$. Yet, in the same conditions in 5 billion years, the velocity
reaches $43$ $500$ $Km$/$s$ and the straight line  segment $111$
$Mpc$.


\quad Galaxy clusters are distributed in space on $Gigaparsec$ scale
lengths and not in isolated pairs as shown in fig. 2. Accordingly,
the illustrative but essential calculation described above has been
repeated for clusters forming three dimensional arrays in space. 
In this case cluster velocities in the range 60000-90000 km/s (see fig. 29 of ref. \cite{codino21book}) are attained by peripheral clusters of the cubic array confirming the crude, oversimplified estimates of two clusters above.


\section{Empirical basis of the electrostatic repulsion of galaxy clusters} 
\quad A solid-rock base of this calculation derives from the
observed gamma-ray fluxes in galaxy clusters. Let us remind that
according to the previous calculation, only 4 protons out of ten thousand remain in the cluster volume
$V_c$,  the rest e.g. 9996 cosmic nuclei out of 10000,  are outside
this volume as summarized by the variable $\xi_A$ and (1. -$\xi_A$)
in $Section$ $3$.

\quad  If cosmic protons and heavier cosmic nuclei massively overflow from
galaxy clusters only a negligible fraction of them will
remain in the nominal cluster volume $V_c$ = (4/3)$\pi$$r_c^3$. On
the contrary electrons reside in the volume $V_c$ albeit slowly
dragged by the electric field $\vec E_c$ toward the cluster
outskirts and, perhaps, sporadically recompacted
to the central cluster zone by indirect gravity action.
Consequently, the gamma-ray flux from the reaction,  \quad p + p
$\rightarrow$ $\pi^0$ + anything $\rightarrow$ $\gamma$ + $\gamma$ +
anything,  \quad is highly suppressed, almost void. Notice that cosmic electrons in cluster volume are routinely
detected in the radio band as a diffuse radio emission (see
$Appendix$ $A.3$ in ref. \cite{codino21book}). Thus, finite and robust
emissivities of diffuse radio emission in clusters around $
10^{-42}$ $erg$/$s$ $cm^3$$Hz$ \cite{murgia} caused by cosmic electrons loudly protrude
face to the inferred scarcity of gamma rays of hadronic origin.

\quad Empirical evidence since almost two decades supports the
absence of gamma-ray fluxes from galaxy clusters against the
predicted rates  \cite{egret}. Based on the upper limits at two
standard deviations of $10^{-9}$ $photons$/$cm^2$/$s$ it was stated
in 2003 : "In conclusion we have to await the first observational
evidence of high-energy gamma-ray emission from galaxy clusters"
\cite{egret}. Presently (2021) this statement is still alive and
more constraining as new sensitive measurements of gamma-ray fluxes
in galaxy clusters \cite{magic, fermi, zandanel, ackermann14, griffin, ackermann16} confirm their paucity and the inferred
scarcity above as well. The recent upper limit at 3 standard deviations is $2.3 \times 10^{-11}$ $photons$/$cm^2$$s$ for energies in the range $0.8$-$100$ $GeV$ \cite{griffin}. Data are summarized by others
\cite{ackermann16} as follows :  "While there is yet no
observational evidence for energetic protons in clusters, the
presence of relativistic electrons is well established from radio
observations" .
In the context of this work the radio emission of cosmic
electrons abounds because cosmic electrons are predominantly
retained in the cluster volume permeated by the cluster magnetic
field and electric fields of individual cluster galaxies while gamma-ray emission from
the reaction, \quad p + p $\rightarrow$ $\pi^0$ + anything
$\rightarrow$ $\gamma$ + $\gamma$ + anything, \quad is highly
suppressed due to the massive overflow of cosmic protons from galaxy
cluster volumes.

%
%
%


\begin{thebibliography}{99}
\bibitem{codino20book} A. Codino, 2020, The ubiquitous mechanism accelerating cosmic rays at all the energies, Societ\`{a} $Editrice$ $Esculapio$, Bologna,  Italy.

\bibitem{codino21book} A. Codino, 2021, On the overflow of cosmic nuclei from galaxy clusters and the expansion of cosmic matter, in press.

\bibitem{murgia} M. Murgia et al., 2009, Comparative analysis of the diffuse radio emission in the galaxy clusters A1835, A2029 and Ophiuchus, A\&A, \textbf{499}, 679-695

  
\bibitem{egret} O. Reimer et al., 2003, EGRET upper limits on the high-energy gamma-ray emission of galaxy clusters, ApJ, \textbf{588}, 155-164.

\bibitem{magic} J. Aleksi\'{c} et al., 2010, Constraining cosmic rays and magnetic fields in the Perseus galaxy cluster with TeV observations by the MAGIC telescopes, ApJ, \textbf{541}, A99 (pp12).

\bibitem{fermi} T. Arlen et al., 2012, Constraints on cosmic rays, magnetic fields and dark matter from gamma-ray observations of the Coma cluster with VERITAS and Fermi, ApJ, \textbf{757}, 123 (14pp).

\bibitem{zandanel} F. Zandanel and S. Ando, 2014, Constraints on diffuse gamma-ray emission from structure formation processes in the COMA cluster, MNRAS, \textbf{440}, 663-671.



\bibitem{ackermann14} M. Ackermann et al., 2014, Search for cosmic-ray-induced gamma-ray emission in galaxy clusters, ApJ, \textbf{787}, 18 (26pp).

\bibitem{griffin} R. D. Griffin et al., 2014, New limits on gamma-ray emission from galaxy clusters, ApJ, \textbf{795}, L21 (5pp).

\bibitem{ackermann16} M. Ackermann et al., 2016, Search for gamma-ray emissions from the Coma clusters with six years of FERMI-LAT data, ApJ, \textbf{819}, 149 (10 pp).





\end{thebibliography}
\end{document}